\documentclass[runningheads,a4paper]{llncs}
\usepackage{url}
\usepackage{threeparttable}
\usepackage{listings}
\usepackage{multirow}
\usepackage{amsmath}
\usepackage{amssymb}
\usepackage{fancybox, graphicx}
\usepackage{url}
\usepackage[table]{xcolor}
\usepackage{textcomp}
\usepackage{lscape}
\usepackage{booktabs} 
\usepackage{colortbl}
\usepackage{paralist}
\usepackage{cite}
\usepackage{eurosym}
\usepackage{graphicx, subfigure}
\begin{document}

\mainmatter  

\title{An n-sided polygonal model to calculate the impact of cyber security events}

\titlerunning{Polygonal Model}

%
%
\author{Gustavo Gonzalez-Granadillo \and Joaquin Garcia-Alfaro \and Herv\'{e} Debar}

%

\institute{Institut Mines-T\'{e}l\'{e}com, T\'{e}l\'{e}com
SudParis, CNRS UMR 5157 SAMOVAR\\ 
9 rue Charles Fourier, 91011 Evry, France\\ 
\{name.last\_name\}@telecom-sudparis.eu}
%
%

\toctitle{RORI}
\tocauthor{RORI}
\maketitle

\begin{abstract}

This paper presents a model to represent graphically the impact of cyber events (e.g., attacks, countermeasures) in a polygonal systems of n-sides. The approach considers information about all entities composing an information system (e.g., users, IP addresses, communication protocols, physical and logical resources, etc.). Every axis is composed of entities that contribute to the execution of the security event. Each entity has an associated weighting factor that measures its contribution using a multi-criteria methodology named CARVER. The graphical representation of cyber events is depicted as straight lines (one dimension) or polygons (two or more dimensions). Geometrical operations are used to compute the size (i.e, length, perimeter, surface area) and thus the impact of each event. As a result, it is possible to identify and compare the magnitude of cyber events. A case study with multiple security events is presented as an illustration on how the model is built and computed.\\


\end{abstract}

\section{Introduction}

A range of difficult issues confront the assessment of the impact of cyber security events \cite{Roberts09}. A set of individual actions performed either by the attacker (e.g., malicious actions executed in order to exploit a system's vulnerability) or by the target system (benign actions executed as a response to an adversary) is hereinafter referred to as a cyber security event.

Computing the economic impact of cyber security events is an open research in the ICT domain. Specialized information security organizations e.g., Computer Emergency Response Team (CERT)\cite{Cert15}, Ponemon Institute\cite{Ponemon15}, Verizon\cite{Verizon15}, etc., perform annual reports on such estimations based on real-world experiences and in-depth interviews with thousands of security professionals around the world. The research is designated to help organizations make the most cost-effective decisions possible in minimizing the greatest risk to their organizations. 

 
Previous researches propose simulation models \cite{Dini13, Dini14} and geometrical models \cite{Gonzal15, GJD15} to estimate and analyze the impact of cyber events. Geometrical models have been the core topic of a variety of research in many disciplines \cite{Liebert10, Emerson11}. However, most of the proposed solutions are limited to three dimensions, making it difficult to provide a graphical representation of geometrical instances in four or more dimensions.

In this paper, we propose a geometrical model to calculate the impact of cyber events in an n-sided polygonal system. The approach considers information about all entities composing an information system (e.g., users, IP addresses, communication protocols, physical and logical resources, etc.), as well as contextual information (e.g., temporal, spacial, historical conditions) to plot cyber attacks and countermeasures as instances of n sides, in a polygonal system. 

In addition, we are able to perform geometrical operations (e.g., length, perimeter, area) over the polygonal instances, which allows us to compare the impact of multiple cyber events. Such comparison provides the means to determine the coverage level i.e., the portion of the incident that is covered by a given security countermeasure and the portion that is left as a residual risk.

The rest of the paper is structured as follows: Section  \ref{pol_sys} presents our proposed polygonal model and discusses about its construction. Section \ref{pol_inst} details the main polygonal instances that result from our model. Section \ref{pol_ope} details the impact measurement of the different geometrical instances. Section \ref{use_case} presents a case study with multiple events (e.g., attacks and countermeasures) to illustrate the applicability of our approach. Related works are presented in Section \ref{rel_work}. Finally, conclusions and perspectives for future work are presented in Section \ref{concl}.

The contributions on this paper are summarized as follows:
\begin{itemize}
\item A geometrical model that projects the impact of security events (e.g., attacks, countermeasures) in an n-sided polygonal system. The instances resulting from the model are straight lines (mono-axial system) or polygons (multi-axial system).

\item A process that performs geometrical operations to calculate the size of the polygonal instances (i.e., length, perimeter, area), which allows us to compare the impact of multiple cyber events.


\item The deployment of our model in a case study with multiple events over several dimensions.

\end{itemize}

\section{Proposed polygonal model}\label{pol_sys}

A polygon is defined as an end to end connected multilateral line which can be expressed as point sequence ($P_0$, $P_1$, $P_2$,..., $P_n$). The $P_0P_1$, $P_1P_2$, ..., $P_{n-1}P_n$ are known as the polygon edges. And the $P_0$,
$P_1$, $P_2$, ..., $P_N$ are referred to as the apex of the polygon\cite{Hai09}. 

Considering the characteristics of access control models \cite{Kalam03, Li06, rbac06}, we identified several entities that contribute directly to the execution of a given attack e.g., User account (subject), Resource (object), and Channel (the way to execute actions, e.g., connect, read, write, etc). In addition, we used the notion of contexts proposed in the Organization based Access Control (OrBAC) model \cite{Cuppens08, Cuppens03}, to extend the approach into an n dimensional system, where every context will be a new dimension, such as information security properties (e.g., confidentiality, integrity, availability); temporal conditions. (e.g., granted privileges only during working hours), spatial conditions (e.g., granted privileges when connected within the company premises), and historical conditions (e.g., granted privileges only if previous instances of the same equivalent events were already conducted).


Our polygonal model is proposed to represent services, attacks and countermeasures as an n-sided polygon, n being the number of entities (e.g, user account, channel, resource, etc). Each entity is projected in one axis of the polygonal system.  There is no limit in the number of axes composing our model. It can be mono-axial (considering only one entity), or multi-axial (considering two or more entities). 


 
Our proposed geometrical model has the following characteristics:

\begin{itemize}
\item There is at least one entity represented in the geometrical instance;
\item The contribution of each entity is represented in one axis of the polygonal system;
\item The contribution of each axis must be greater than zero and no more than one hundred percent;
\item The end points of the instance axes are connected to form a polygon;
\item The union of two end points represents one side of the polygon;
\item Polygons can be regular, irregular, and/or convex;
\item Concave polygons are excluded from our model since it is not possible to plot instances in which one or more interior angles are greater than 180 degrees;
\item Polygons are closed with no holes inside;
\item Polygons are not self-intersecting;
\end{itemize}
 
The remaining of this section gives examples of the possible entities that can be used to calculate the impact of cyber events and details the contribution calculation of each side of our polygonal model.

\subsection{Entities of the Polygonal Model}

An entity is an instance that exists either physically or logically. Entities regroup elements with similar characteristics or properties. An entity may be a physical object such as a house or a car (they exist physically), an event, such as a house sale or a car service, or a concept such as a customer transaction or order (they exist logically as a concept). Examples of entities used in our polygonal model are given as follows:

\subsubsection{User Account} It considers all active user accounts from the system. A user account is the equivalent of a ‘subject’ in an access control policy. User accounts are associated to a given status in the system, from which their privileges and rights are derived (e.g., system administrator, standard user, guest, internal user, nobody).

\subsubsection{Resource} It considers physical components (e.g., host, server, printer) and logical components (e.g., files, records, database) of limited availability within a computer system. A resource is the equivalent of an ‘object’ in an access control policy. 

\subsubsection{Channel} In order to have access to a particular resource, a user must use a given channel. A channel is the equivalent to an ‘action’ in an access control policy. We consider the IP address and the port number to represent channels in TCP/IP connections. However, each organization must define the way its users connect to the system and have access to the organization's resources. \\

Other entities can consider temporal conditions (e.g., connection time, detection time, time to react, time to completely mitigate the attack, recovery time, etc.), spatial conditions (e.g., user's location, security areas, specific buildings, a country, a network or sub-network, etc.)

In addition, an event can be associated to a particular issue compromising the system's confidentiality (e.g., unauthorized access to sensitive information, disclosure resources, etc), integrity (e.g., unauthorized change
of the data contents or properties, etc), or availability (e.g., unavailable resources, denial of service, etc). 

Every organization must define their own entities based on their historical data, expert knowledge and assessments they perform on their systems. 

\subsection{Dimension Contribution}\label{dim_cont}

Each side contributes differently in the impact calculation of the polygon. This contribution represents the affectation of a given element in the execution of an event. Following the CARVER methodology \cite{Norman10, Fed91}, which considers six criteria (i.e., criticality, accessibility, recuperability, vulnerability, effect, recognizability), we assign numerical values on a scale of 1 to 10 to each type of element within the axis. As a result, we obtain a weighting factor (WF) that is associated to each type of element. Examples of the practical implementation of this methodology in real case scenarios can be seen in \cite{Gonzal15, GJD15}.

The contribution $Co$ of each side $D$ in the execution of an event $E$ is a value than ranges from zero (when there is no element of the dimension affected to a given event), to one (when all elements of the dimension are affected to a given event).  The contribution of a side $D$ is calculated using Equation \ref{eq:cont}.

\begin{equation}\label{eq:cont}
Co(D,E)=\frac{\sum_{j=1}^{n}Y_j \times WF(Y_j) \hspace{2mm} \forall j \in Y}{\sum_{i=1}^{n}X_i \times WF(X_i) \hspace{2mm} \forall i \in X}
\end{equation}

\noindent Where\\
\noindent $X$ = total number of elements\\
\noindent $Y$ = affected elements\\
\noindent $WF$ = Weighting Factor\\

In order to apply Equation \ref{eq:cont} in a practical case, let us consider the axis defined in the previous section. The contribution for the user account dimension, for instance, can be evaluated as the number of users affected by a given attack over the total number of active users from the system. Similarly, the contribution of the confidentiality dimension can be evaluated as the number of alerts indicating a confidentiality issue over the total number of alerts in a given period of time. For spacial contexts we can evaluate the number of incidents occurring in a given location over the total number of reported incidents within a period of time.

\section{Resulting Polygonal Instances}\label{pol_inst}
A variety of geometrical instances (e.g., regular and irregular polygons such as: line segments, triangles, squares, pentagons, etc.) results from the analysis of the entities' information included in a system, attack and/or countermeasure. By definition, polygons are not allowed to have holes in them [23]. The remaining of this section details the different polygonal instances.

\subsection{One Dimension}
\begin{figure}[ht!]
     \begin{center}
        \subfigure[Straight Line]{
            \label{fig:dim1}
            \includegraphics[width=0.44\textwidth]{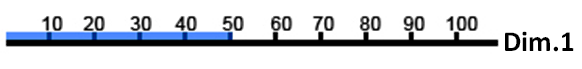}
        }
        \subfigure[Right Isosceles Triangle]{
            \label{fig:dim2}
            \includegraphics[width=0.44\textwidth]{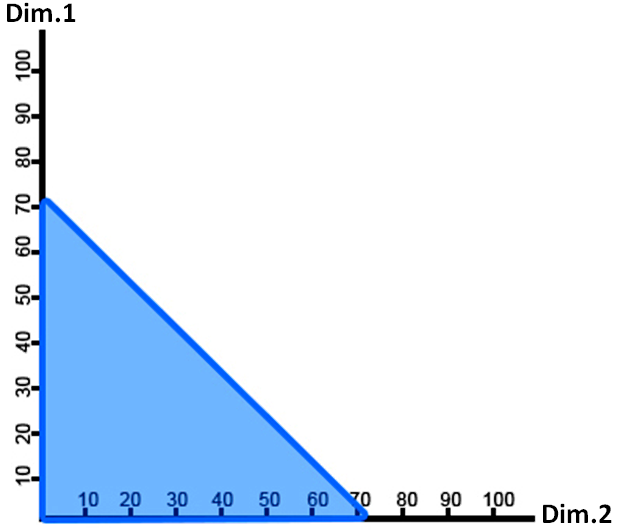}
        }
        \end{center}
    \caption{Impact graphical representation in one and two dimensions}
   \label{fig:test}
\end{figure}

Plotting the contribution of one dimension into our polygonal system results into a line segment. Let us consider, for instance, an attack $A_1$ that compromises standard users $U1:U5$ ($WF=2$) and admin $U11:U20$ ($WF=5$), from a list of 30 users (users $U1:U10$ with $WF=2$ and users $U11:U30$ with $WF=5$). The contribution of this dimension will be equal to $Co(Dim1)$ = $\frac{(5 \times 2)+(10 \times 5)}{(10 \times 2)+(20 \times 5)}$ = $0.5$. Figure \ref{fig:dim1} shows the graphical representation of the impact contribution of attack $A_1$ over the user dimension ($Dim_1$). 

\subsection{Two Dimensions}
When we have information of two dimensions of our polygonal system (e.g., resources and channels, or users and location), we plot the information to obtain polygons in two dimensions (i.e., right triangles). For instance, an attack that compromises 70\% of resources ($Dim_1$), using 70\% of the system's channels ($Dim_2$), will be represented as a right isosceles triangle\footnote{Triangle with a right angle and two equal sides and angles} (Figure \ref{fig:dim2}); the same attack that compromises 40\% of resources ($Dim_1$), using 70\% of the system's channels ($Dim_2$) will be represented as a right scalene triangle\footnote{Triangle with a right angle and all sides of different lengths}.

\subsection{Three Dimensions}
\begin{figure}[ht!]
     \begin{center}
        \subfigure[Scalene Triangle]{
            \label{fig:dim3}
            \includegraphics[width=0.44\textwidth]{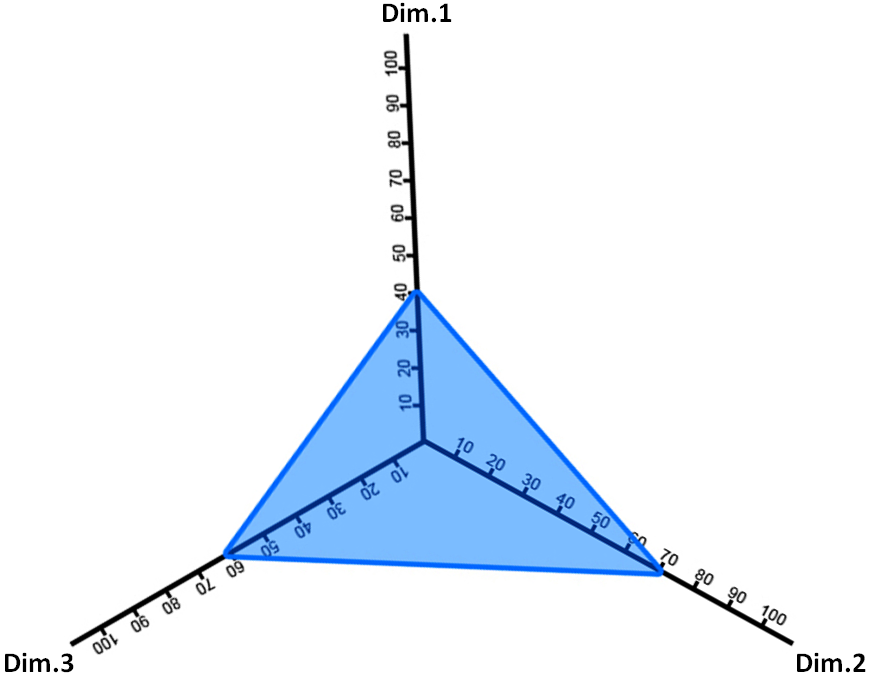}
        }
        \subfigure[Rhomboid]{
            \label{fig:dim4}
            \includegraphics[width=0.44\textwidth]{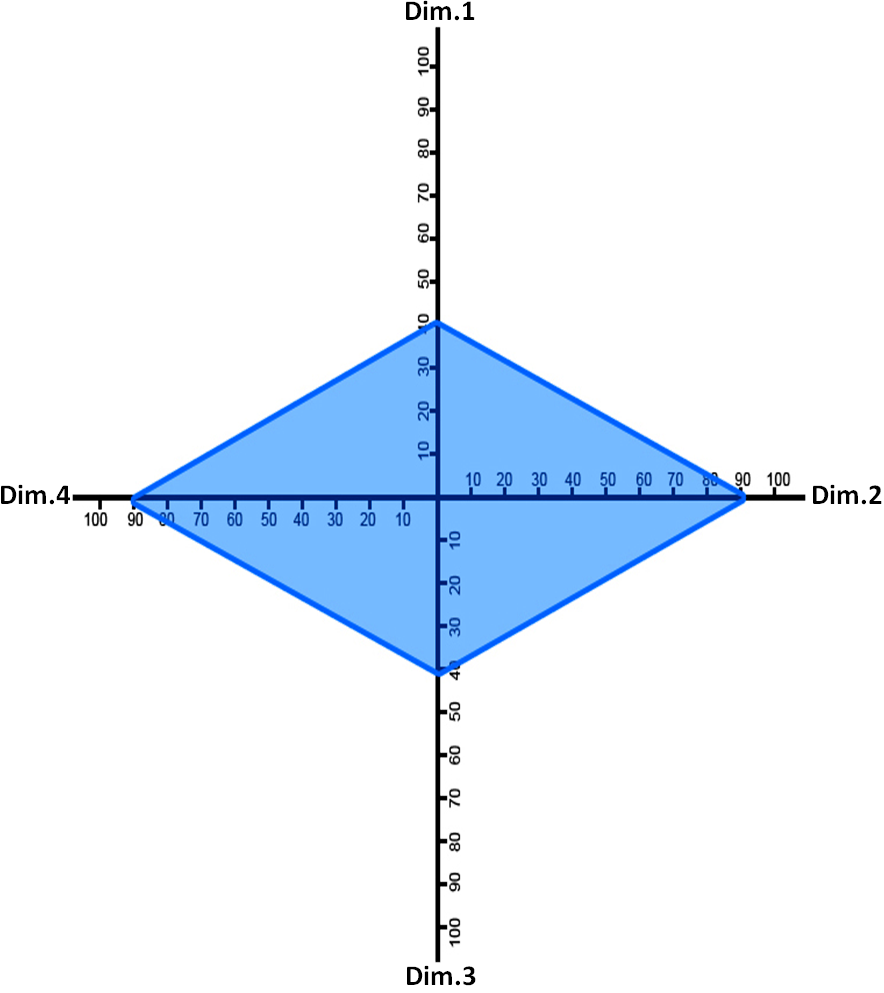}
        }
        \end{center}
    \caption{Impact graphical representation in three and four dimensions}
   \label{fig:test}
\end{figure}

The representation of the impact contribution in three dimensions results into any type of triangles except for right triangles. For instance, an attack with 70\% of resources, users and channels contribution will be represented as an equilateral triangle\footnote{Triangle in which all three sides are equal and all three internal angles are congruent to each other}; the same attack with 40\% of resource contribution, 70\% of user contribution, and 60\% of channel contribution will be graphically represented as a scalene triangle\footnote{Triangle with all sides and angles unequal} (Figure \ref{fig:dim3}). 

\subsection{Four Dimensions}
Four-dimensional geometry is Euclidean geometry extended into one additional dimension. The graphical representation of the impact contribution of a given event in four dimensions results into a quadrilateral\footnote{Polygon with four sides and vertices (e.g., square, rhombus, kite, etc)} . We discard rectangles, since it is not possible to represent instances that have both: two equal alternate sides and right angles. In addition, we discard rhombus from our graphical representation, since it is not possible to represent instances that have both: equal lengths and non-right angles. 

Let us consider, for instance, an attack with 40\% of contribution in four dimensions: resources ($Dim_1$) users ($Dim_2$), channels ($Dim_3$), and recovery time ($Dim_4$) will be represented as a square. The same attack compromising 40\% of resources ($Dim_1$) and channels ($Dim_3$), 10\% of users ($Dim_2$), and 70\% of the recovery time ($Dim_4$) will be graphically represented as a kite\footnote{Quadrilateral whose four sides can be grouped into two pairs of equal-length sides that are adjacent to each other}. Similarly, the same attack compromising 40\% of resources ($Dim_1$) and channels ($Dim_3$), and 90\% of users ($Dim_2$) and recovery time ($Dim_4$) will be represented as a rombhoid\footnote{Parallelogram in which adjacent sides are of unequal lengths and angles are non-right angled}, as shown in Figure \ref{fig:dim4}.

\subsection{N Dimensions}

Following the same approach as in previous examples, we propose to represent the impact of each dimension composing our polygonal system as segments, and to connect them to form a 2D (regular or irregular) closed polygon (e.g. pentagon, hexagon, octagon, etc.)

\begin{figure}[ht!]
     \begin{center}
        \subfigure[Irregular Pentagon]{
            \label{fig:dim5}
            \includegraphics[width=0.44\textwidth]{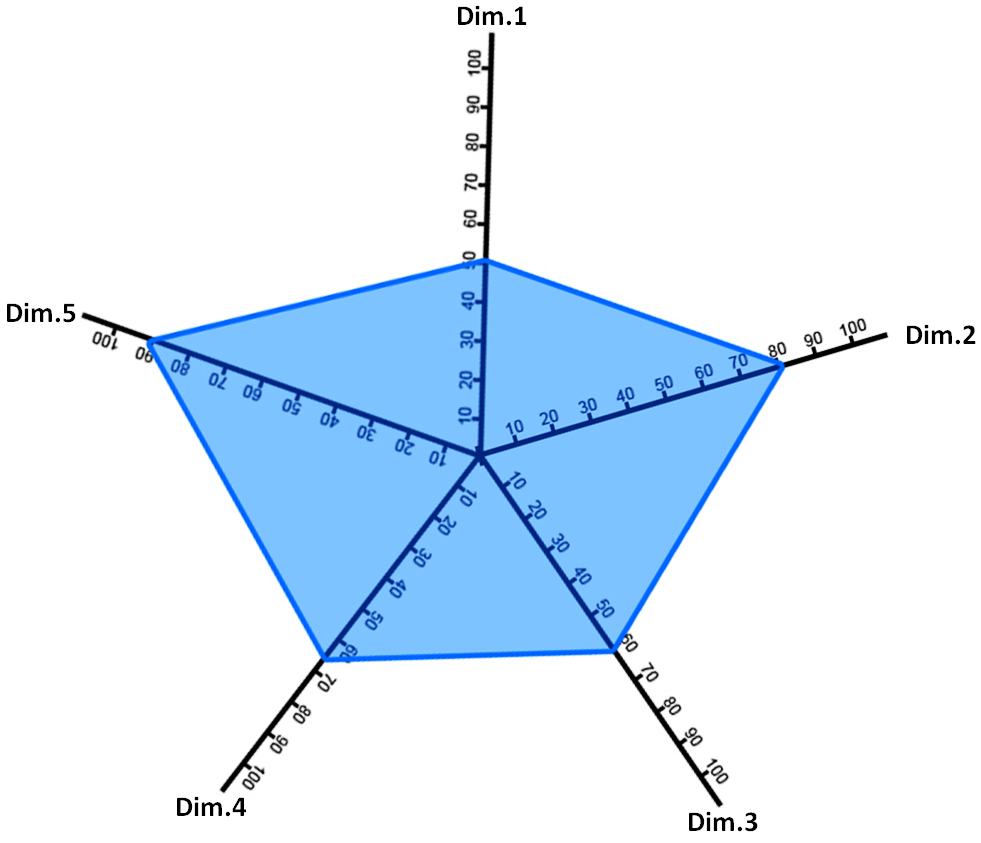}
        }
        \subfigure[Irregular Octagon]{
            \label{fig:dim8}
            \includegraphics[width=0.44\textwidth]{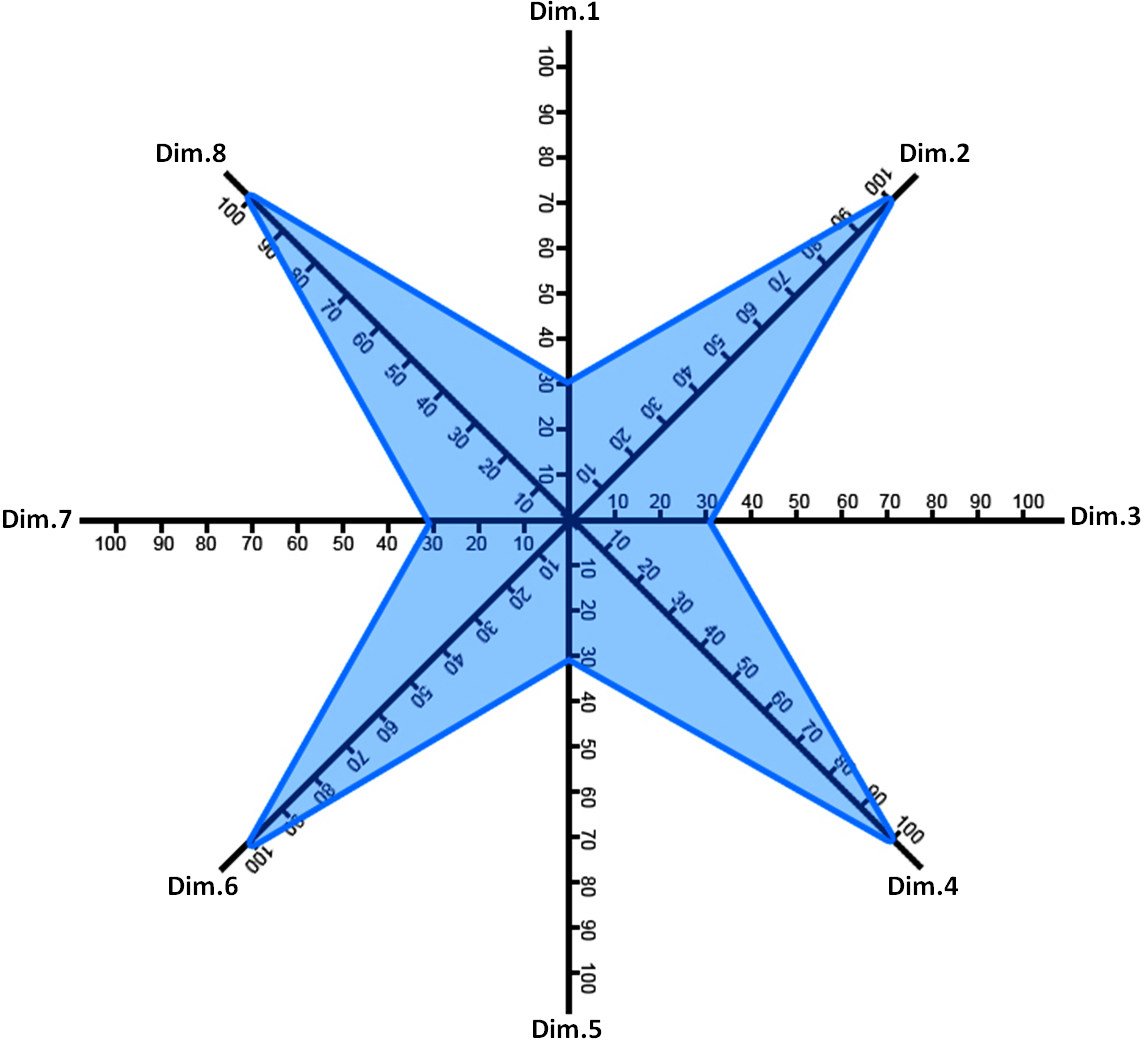}
        }
        \end{center}
    \caption{Impact graphical representation in more than four dimensions}
   \label{fig:test}
\end{figure}

 For instance, let us assume that we have information of attack $A_1$ affecting five dimensions:  $Co(Dim_ 1)=50\%$, $Co(Dim_2)=80\%$, $Co(Dim_3)=60\%$, $Co(Dim_4)=65\%$, and $Co(Dim_5)=90\%$. The contribution impact of attack $A_1$ is graphically represented as an irregular pentagon, as depicted in Figure \ref{fig:dim5}. 


The model selects all elements affected in each dimension to represent it as a continuous segment that indicates the impact of such dimension for that particular event. We connect them all in order to form an n-polygon (n being the number of dimensions of the polygonal system). 

In addition, Figure \ref{fig:dim8} depicts the graphical representation of an irregular octagon, where  the contribution of the odd dimensions is $30\%$, and the contribution of the even dimensions is $100\%$. A variant of this case will be if the contribution to an event on one or more dimensions is zero. In such a case, the dimension will be discarded from the graphical representation.

\section{Geometrical Operations}\label{pol_ope}
This section details the measurements of the different geometrical figures described in the previous section. Such measurement allows the mathematical computation of the impact of multiple events in the system. 

\subsection{Length of Polygons}
The length of a straight line corresponds to the distance from its origin to its endpoint. In a mono-axial polygonal system, the length is computed as the impact contribution of such dimension over the event. Results are expressed in units, using Equation \ref{eq:cont}. In a bi-axial or multi-axial polygonal system, the length is the equivalent of the perimeter of a polygon. 

The perimeter of a regular polygon equals the sum of the lengths of its edges. A regular polygon may be defined by the number of its sides $n$ and by its radius $R$, that is to say, the constant distance between its center and each of its vertex. The perimeter of a regular polygon is computed using Equation \ref{eq:reg_pol}.

\begin{equation}\label{eq:reg_pol}
P(Regular Polygon) = 2 \times n\times R\times sin(\frac{180}{n})
\end{equation}

Particular cases can be defined. In a bi-axial polygonal system, for instance, the perimeter ($P$) of an event is calculated as the sum of the impact contribution of each dimension to the event and the length of the connecting side of the two axes, as shown in Equation 1. For equilateral polygons (e.g., hexagon, heptagon,...) whose edge's length is known, we calculate the perimeter using Equation 2, whereas for irregular polygons, we use Equation 3 to calculate their perimeter.
\small
\begin{eqnarray}
P(Right Triangle)=& Co(Dim_1)+Co(Dim_2)+ L(X)\\
P(Equilateral Polygon)=& n \times L(X)\\
P(Irregular Polygon) =& \sum_{i=1}^{n} L_i(X)
\end{eqnarray}

\normalsize

\noindent Where\\
\noindent $L$, $L_1$, $L_2$,..., $L_n$ = length of the edges of the polygon\\
\noindent $n$ = Number of sides of a regular polygon

Let us consider, for instance, a regular polygon of five sides (i.e., pentagon), with each dimension contribution equals to $10\%$. The perimeter of the pentagon is calculated as $P(Regular Pentagon)$ = $2\times 5\times 10 \times \sin(45)$= $58.78$ $units$. For irregular polygons, the perimeter is calculated as the sum of the length of each edge (Equation 3), considering the same pentagon, whose edges measure (10, 25, 10, 45, 20), the perimeter of such polygon is equal to $P(Iregular Pentagon)$ = $110$ $units$.

\subsection{Area of Polygons}
The area (A) of a given event measures the amount of space inside the boundary of a flat (2-dimensional) object such as a triangle or square. 

For regular polygons, the area equals the product of the perimeter and the apothem\footnote{The line segment from the center of a regular polygon to the midpoint of a side} divided by two. Results are expressed in $units^2$, using Equation \ref{eq:area_reg}.

\begin{equation}\label{eq:area_reg}
A (Regular Polygon) = \frac{Perimeter \times Apothem}{2}
\end{equation}

For irregular polygons, we compute the area as the sum of the contribution value of $Dim_i$ times the contribution value of $Dim_{i+1}$ divided by two, as shown in Equation \ref{eq:area_irreg}.

\begin{equation}\label{eq:area_irreg}
A (Irregular Polygon) = \frac{\sum_{i=1}^{n}Co(Dim_i)\times Co(Dim_{i+1})}{2}
\end{equation}


For the previous Equation, note that in the last term (i.e., $Co(Dim_n)$), the expression must wrap around back to the first term (i.e., $Co(Dim_1)$). This method works correctly for triangles, regular and irregular polygons, as well as convex and concave polygons, but it will produce wrong answers for self-intersecting polygons, where one side crosses over another. However, such cases are excluded from our research.

Let us take an example of an attack $A_1$ that affects 60\% of resources ($Dim_1$), 60\% of channels ($Dim_2$), 80\% of users ($Dim_3$) and requires 40\% of recovery time ($Dim_4$). Attack A1 will have an area equal to $A(Quadrilateral)$ = $[(60\times 60)$ + $(60\times 80)$ + $(80\times 40)$ + $(40\times 60)]/2$ = $700$ $units^2$

\section{Case Study: }\label{use_case}

A vulnerability in OpenSSH (i.e., CVE-2015-5600) has been exploited to bypass the maximum number of authentication attempts and launch attack $A_1$ (brute force attack against a targeted server).  The vulnerability is related to the keyboard-interactive authentication mechanism and it can be exploited through the KbdInteractiveDevices option. The crucial part is that if the attacker requests 10,000 keyboard-interactive devices, OpenSSH will gracefully execute the request and will be inside a loop to accept passwords until the specified devices are exceeded.
A remote attacker could therefore try up to 10,000 different passwords and they would only be limited by a login grace time setting, which by default is set to two minutes. Attack $A_1$ affects a great number of users, channels, resources, and systems where keyboard-interactive authentication is enabled. Three security countermeasures have been proposed to mitigate attack $A_1$. Table \ref{Table:dim_info} summarizes this information.\\

\noindent {\bf Proposed Countermasures:}
\begin{enumerate}[{C.}1]
\item Install an OpenSSH patch 
\item Limit access to SSH in the firewall,
\item Disable password authentication for the root account
\end{enumerate}

\begin{table} [h]
\caption{Events Dimensional Information} 
  \label{Table:dim_info}
\begin{center}
\begin{tabular}{p{2.3cm}p{2.3cm}p{1cm}p{1cm}p{1cm}p{1cm}p{1cm}p{1cm}}
\hline
\hline	
{\bf Dimension}&{\bf Category}& {\bf Q}& {\bf WF}& {\bf A1}& {\bf C1}& {\bf C2}& {\bf C3}\\
\hline
Internal  & root &3& 5 &3 & 3& 3& 3\\
  user    & standard user &25& 2 &25 & 25& 25& -\\ 
\hline
Channels & credentials &28& 4 &28 & -& -& 3\\
         & IP addresses &30& 3 &- & -& 30& -\\ 
\hline
Physical  & PC &27& 2 &- & 27& -& -\\
resources & server &12& 5 &5 & 3& -& 12\\ 
\hline
Logical  & firewall &2& 4 &2 & -& 2& -\\
resources & software &10& 3 &4 & 4& -& 5\\ 
 
\hline
\hline

\end{tabular}
\end{center}
\end{table}

The first two columns from Table \ref{Table:dim_info} identify the four main dimensions and categories of each dimensions respectively. The next two columns shows the number of elements (Q) composing each category of the dimension, and their corresponding weighting factor (WF). The rest of the columns show the number of elements affected by attack $A_1$ and countermeasures $C_1$, $C_2$, and $C_3$.  

\subsection{Impact Calculation}

1. System Dimensions: We compute the system's dimensions using information from Table \ref{Table:dim_info}, as follows:
\begin{itemize}
\item Internal User (IU) = $(3 \times 5) + (25 \times 2)$ = $65$ $units$
\item Channels (Ch) = $(28 \times 4) + (30 \times 3)$ = $202$ $units$
\item Physical Resources (PR) = $(27 \times 2) + (12 \times 5)$ = $114$ $units$
\item Logical Resources (LR) = $(2 \times 4) + (10 \times 3)$ = $38$ $units$
\end{itemize}

2. Dimension Contribution: We calculate the contribution of each dimension on the execution of the events (i.e.,$A_1$, $C_1$, $C_2$, and $C_3$) with respect to the system value, using Equation \ref{eq:cont}. For instance, Attack $A_1$ affects the following dimensions: 
\begin{itemize}
\item IU = $(3 \times 5) + (25 \times 2)$ = $65$ $units$ $\rightarrow$ $100\%$
\item Ch = $28 \times 4$ = $112$ $units$ $\rightarrow$ $55.45\%$
\item PR = $5 \times 5$ = $25$ $units$ $\rightarrow$ $21.93\%$
\item LR = $(2 \times 4) + (4 \times 3)$ = $20$ $units$ $\rightarrow$ $52.63\%$
\end{itemize} 

3. Impact Calculation: We calculate the geometrical operations of attack $A_1$ and countermeasures $C_1$, $C_2$, and $C_3$, using Equations 5 and \ref{eq:area_irreg}. For attack $A_1$, for instance, we compute the perimeter and area as follows:

\small
\noindent P($A_1$) = $L(IU-Ch)$ + $L(Ch-PR)$ + $L(PR-LR)$ + $L(LR-IU)$
\normalsize
\noindent P($A_1$) = $114.34$ + $59.62$ + $57.02$ + $113.00$ = $343.99$ $units$

\noindent A($A_1$) = $\frac{(100\times 55.45)+ (55.45 \times 21.93)+ (21.93\times 52.63)+ (52.63\times 100)}{2}$

\noindent A($A_1$) = $6,588.91$ $units^2$







Table \ref{Table:impact_info} summarizes the impact values of all events.

\begin{table} [h]
\caption{Event Impact Evaluation} 
  \label{Table:impact_info}
\begin{center}
\begin{tabular}{p{4cm}p{3cm}p{3cm}}
\hline
\hline	
{\bf Event}&{\bf P($units$)}& {\bf A($units^2$)}\\
\hline
$S$&565.69 &20,000.00\\
$A_1$  & 343.99 &6,588.91 \\
$C_1$  & 333.66 &2,534.63\\
$C_2$  & 277.28 &3,280.35 \\
$C_3$  & 188.31 &1,719.12\\
$C_1 \cup C_2$  & 357.77 &6,110.71 \\
$C_1 \cup C_3$  & 340.76 &3,645.09\\
$C_2 \cup C_3$  & 351.73 &6,412.67 \\
$C_1 \cup C_2 \cup C_3$  & 364.40 &6,744.36\\

\hline
\hline

\end{tabular}
\end{center}
\end{table}

As depicted in Table \ref{Table:impact_info}, attack $A_1$ is compared against the system $S$ and countermeasures $C_1$, $C_2$, and $C_3$. Attack $A_1$ affects $32.94\%$ of the total system area. Applying countermeasures individually will reduce part of the attack impact. However, if multiple countermeasures are implemented, the risk is expected to be reduced substantially. The best solution for this attack scenario is implementing $C_2$ and $C_3$, since the application of the three countermeasures will probably increase costs and potential collateral damages with no improvement in the mitigation level of attack $A_1$.

\subsection{Graphical Representation}
\begin{figure}[ht!]
    \begin{center}
            \includegraphics[width=0.75\textwidth]{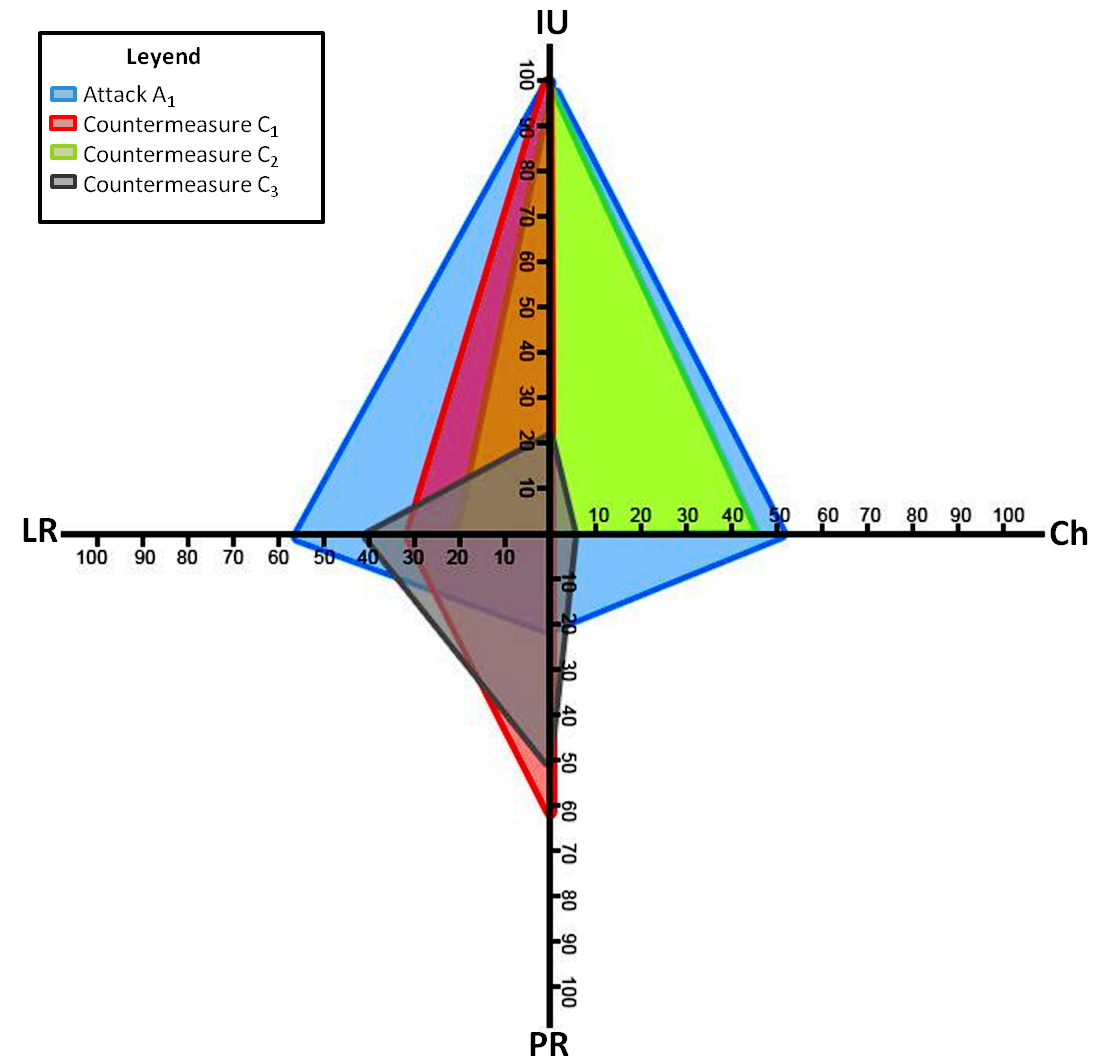}
        \end{center}
    \caption{Impact graphical representation of events - Case by Case Analysis}
   \label{fig:007}
\end{figure}

Figure \ref{fig:007} shows the graphical representation of attack $A_1$ (in blue) and the individual implementation of countermeasures $C_1$ (in red), $C_2$ (in green), and $C_3$ (in grey). The graphical representation shows a case by case implementation of the different security countermeasures. It is important to note that each countermeasure affects a given set of elements in at least one dimension. Countermeasure $C_2$, for instance, only affects elements that are vulnerable to attack $A_1$, whereas Countermeasures $C_1$, and $C_3$ requires modifications of elements that are not vulnerable to attack $A_1$ (e.g., physical resources). 

The visualization of cyber attacks and countermeasures in the same geometrical space helps security administrators in the analysis, evaluation and selection of security actions as a response to cyber attacks. It is possible to identify priority areas (e.g., those where most attacks are concentrated, or where more elements of the system are vulnerable), and perform reaction strategies accordingly. It is also possible to visualize the portion of the attack (e.g., the area of the polygon) that is being controlled by a security countermeasure, and the portion that is left with no treatment (e.g., residual risk). 

\begin{figure}[ht!]
     \begin{center}
            \includegraphics[width=0.75\textwidth]{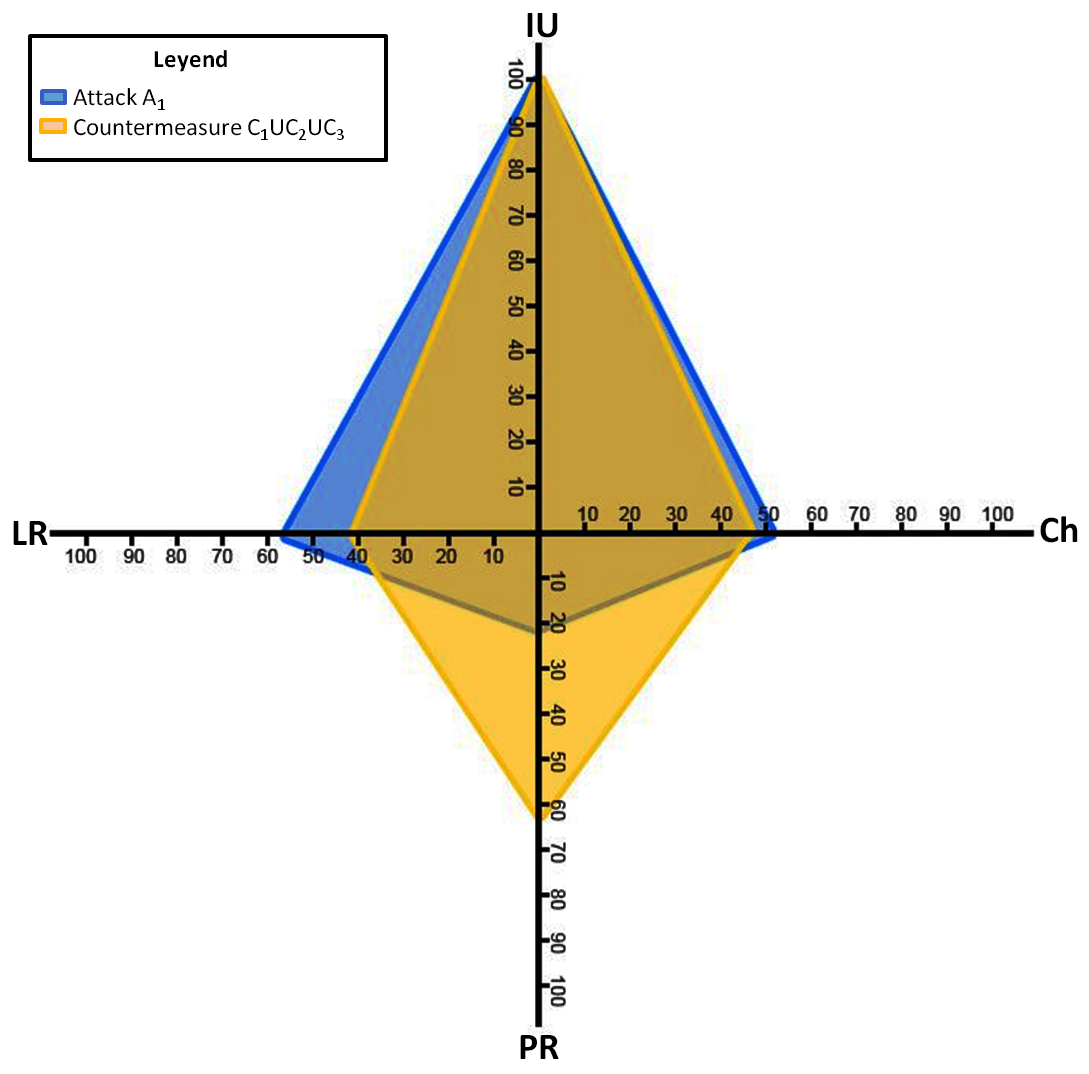}
       \end{center}
    \caption{Impact graphical representation of events - Combined Analysis}
   \label{fig:008}
\end{figure}

Furthermore, it is also possible to plot multiple cyber attacks occurring simultaneously in the system. The same can be performed for multiple countermeasures that need to be implemented simultaneously. The graphical representation of the resulting instance will generally cover a grater area than their individual representations. For instance, the graphical representation of the three countermeasures implemented simultaneously is depicted in Figure \ref{fig:008}, where attack $A_1$ is represented by the blue polygon, and the set of countermeasures is represented by the yellow polygon. 

For this example, the implementation of multiple countermeasures increases the coverage area of the attack, which in turn reduces the attack impact, making it look more attractive than their individual implementation.

\section{Related Work}\label{rel_work}

Determining the impact of cyber security events is an open research in the ICT domain. Several research works rely on metrics to quantitatively measure such impacts. Howard et al., \cite{Howard04, Howard05} and Manadhata et. al. \cite{Manadhata06, Manadhata08, Manadhata10}, for instance, propose a model to systematically measure the attack surface of different software. However, the approach presents the following shortcomings: it cannot be applied in the absence of source code; it includes only technical impact; it cannot be used to compare different system environments; and it does not evaluate the impact of multiple attacks occurring simultaneously in the system.

Other researchers rely on simulations to analyze and estimate the impact of cyber events. Dini et. al \cite{Dini13, Dini14}, for instance, present a simulative approach to attack impact analysis that allows for evaluating the effects of attacks, ranking them according to their severity, and provides valuable insights on the attack impact since during the design phase. The research differs from our work, since their simulation does not provide quantitative analysis on the impact of countermeasures while evaluating the impact of attacks.

Other approaches use geometrical models to provide a 3D view of the events in a variety of disciplines. Emerson et al. \cite{Emerson11}, for instance, propose a geometrical 3D model for use within sport injury studies in order to influence the design of sport equipment and surfaces, which could help to prevent sports injuries. In addition, Liebel and Smitch \cite{Liebert10}, present a geometrical approach for multi-view object class detection that allows performing approximate 3D pose estimation for generic object classes. However, geometrical models are limited to a 3D projection.

2D models have been proposed in a variety of domains \cite{Gao10, Ansari11, Sommer15, Zhang16} as synthetic and generic visualization models that overcome previous drawbacks from 3D representations. 2D models enable viewers to visualize the overall big picture and the interrelationships of various entities. Users may be able to observe how the changes on selected events could potentially affect the overall system to provide understanding on interrelated impacts. However, since the  model provides an abstract picture of one or multiple events, its visualization does not provide an accurate value of the impact coverage (e.g., it is not possible to identify the exact mitigation level of a given countermeasures). It is therefore important to combine the visualization approach with geometrical operations that quantitatively indicate the level at which an attack is controlled by a mitigation action.    


\section{Conclusions}\label{concl}

Based on the limitations of the current solutions, we propose a geometrical approach to project the impact of cyber events in an n-dimensional polygonal system. The approach uses geometrical operations to compute the size of the polygon, and thus the impact of the represented event. As a result, we are able to project multiple events (e.g., attacks, countermeasures), in a variety of axes (e.g., users, channels, resources, CIA, time, etc.), which provides the means to propose security countermeasures as a reaction strategy to mitigate the detected attacks. 

The main novelty of the approach is the use of multiple criteria to build the n-sided polygon using the dimension contribution Equation discussed in Section \ref{dim_cont} and the use of metrics (i.e., length and area), as discussed in Section \ref{pol_ope}. As such, we overcome previous drawbacks about visualization (e.g., inability to plot the impact of cyber security events in four or more dimensions). Results show that implementing multiple countermeasures simultaneously increases the protection area and thus reduces the impact of a given attack. 

Future work will concentrate in quantifying the residual risk and potential collateral damage that result out of the implementation of a set of countermeasures. In addition, we will include other event's related information (e.g., attackers knowledge, capabilities, etc) in order to explore external dimensions that could influence in the impact calculation of a cyber security event.

\section*{Acknowledgment}
The research in this paper has received funding from PANOPTESEC project, as part of the Seventh Framework Programme (FP7) of the European Commission (GA 610416).

\bibliographystyle{abbrv}
\bibliography{sigproc}  

\end{document}